\documentclass[12pt]{iopart}
\usepackage{iopams}  
\usepackage{graphicx}
\usepackage{color}
\usepackage{caption}
\usepackage{subcaption}
\usepackage[dvipsnames]{xcolor}
\usepackage{psfrag,epsfig}
\usepackage[percent]{overpic}
\usepackage{float}
\usepackage{color}
\usepackage{multirow}

\usepackage{tikz}
\usetikzlibrary{arrows.meta,bending,positioning,intersections}
\usepackage{pgfplots}
\usepackage[all]{xy}

\begin{document}

\title[Oscillatory instability in an Ostwald ripening process]{
Oscillatory instability in an Ostwald ripening process}

\author{Michael Wilkinson}

\address{
School of Mathematics and Statistics,
The Open University, Walton Hall, Milton Keynes, MK7 6AA, United Kingdom
}
\vspace{10pt}
\ead{
m.wilkinson@open.ac.uk
}
\begin{indented}
\item March 2025
\end{indented}

\begin{abstract}
This paper considers an Ostwald ripening process
in which new droplets are injected at a constant rate,
with a fixed distribution of radii, and in which droplets are removed 
when they grow to a specified maximum radius. 
This process exhibits a transition from a steady state 
to a limit cycle as a parameter is varied. 
The instability is shown to be related to the 
roots of the Laplace transform of a response kernel.
A model is described which gives a good approximation of the period of the 
limit cycle. The model may also exhibit chaotic behaviour. 
The relevance of the model to atmospheric precipitation is discussed.
\end{abstract}
\maketitle

\section{Introduction}
\label{sec: 1}

Ostwald ripening (reviewed in \cite{Voo84}) is a classic problem in statistical physics, which has 
been intensively studied. The original context was annealing of metal alloys
\cite{Ost96}, but in this work it will be discussed in the simpler context of a dispersion 
of liquid droplets in a gas phase. The physical process, which will be fully 
explained in section \ref{sec: 2}, is an evolution of the droplet size distribution 
which results in smaller droplets undergoing evaporation as a consequence 
of the their higher Laplace pressure. Material which evaporates from the  smaller droplets
condenses upon the larger ones. An analysis by Lifshitz and Slyozov \cite{Lif+58} (see also
\cite{Lif+61,Wag61}) is the basis of most theoretical discussions, and its predictions 
are in quite good agreement with experimental observations (although there are
systematic deviations \cite{Voo84}).

In order to analyse Ostwald ripening, it is necessary to consider 
the dynamics of both the droplet sizes, and the supersaturation.
The approach in \cite{Lif+58} concentrates on the dynamics 
of the droplet sizes, and the long-time evolution of the supersaturation is deduced 
by requiring that the droplet size distribution does not develop 
unphysical characteristics. It is often argued that the supersaturation 
field contains a negligible fraction of the material which forms 
the droplet phase, leading to a prediction 
that the critical radius is equal to the mean radius of the surviving particles
(see, e.g. \cite{Nie+99}). Penrose \cite{Pen97} suggested a way 
to incorporate information about the supersaturation field by relating 
the macroscopic phenomenon of Ostwald ripening to the microscopic 
Becker-D\"oring process, which leads to an equivalent formulation. 

This paper will consider the macroscopic 
evolution equation for the supersaturation explicitly. 
One motivation was to arrive at 
a clearer understanding of the dynamics of the supersaturation.
 The equations of motion for the system are discussed in 
 section \ref{sec: 2}, where it is shown that the equation of 
 motion for the supersaturation contains a dimensionless parameter.
 Section \ref{sec: 3} discusses the estimate of the dimensionless parameter 
for the atmospheric aerosol, and goes on to consider the consequences
of its large value, both in general terms, and in the context of determining the 
stability of a steady-state solution. It is argued that the large value of the dimensionless
parameter supports the suggestion that the critical radius is equal to the mean 
radius of the surviving particles, consistent with the approximations considered 
in \cite{Nie+99,Pen97}.

Another objective of this work was to investigate whether 
the large parameter in the equation of motion for supersaturation 
might be associated with instability of Ostwald ripening processes. 
It is easier to assess and characterise instability if the system 
has a steady state. The classic version of Ostwald ripening
does not exhibit a conventional steady state, because the number of droplets 
decreases as a function of time. This paper considers an alternative 
model, in which small droplets are continually introduced (at a 
steady rate, and with a fixed distribution of sizes). Droplets either 
evaporate, or else grow in radius, until they are removed above 
a certain critical radius. Section \ref{sec: 3} discusses a criterion for the stability of this 
model, expressed in terms of the Laplace transform of a response kernel.

This model could be relevant in atmospheric 
physics, where the supply of nucleation centres for condensation 
of droplets is continuously being replenished, and where large 
droplets are lost because they fall out of suspension due to gravity.
The model is also relevant to experiments in which a binary mixture 
of liquids undergoes phase separation when the temperature is slowly 
driven through a critical point. These systems can undergo periodic
variations of the turbidity due to coarsening of the separating phases 
\cite{Vol+97}. These oscillations of turbidity can be explained by mechanisms
involving Ostwald ripening \cite{Wil14}.

\begin{figure}
\begin{center}
\includegraphics[width=5.5cm]{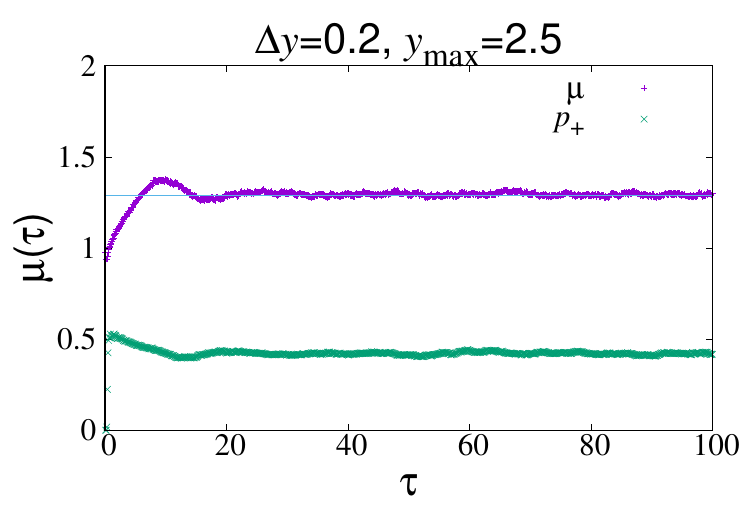}\includegraphics[width=5.5cm]{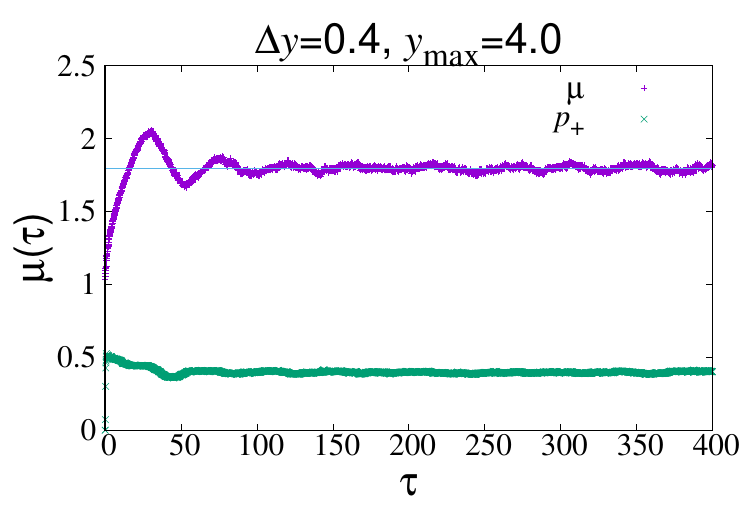}
\includegraphics[width=5.5cm]{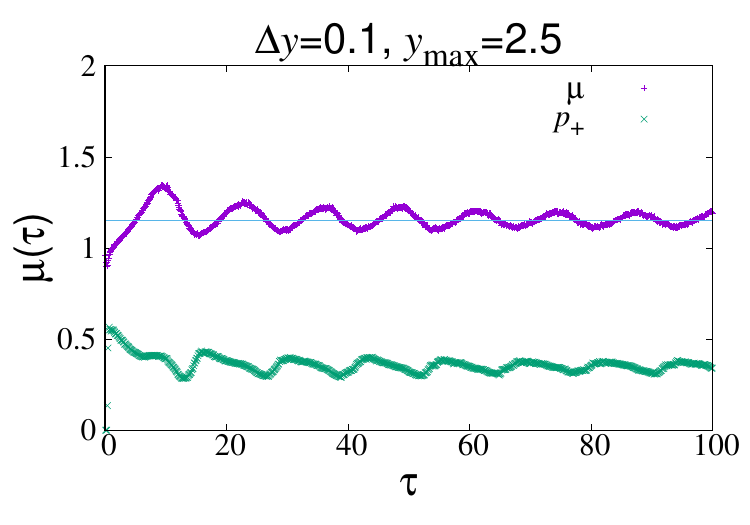}\includegraphics[width=5.5cm]{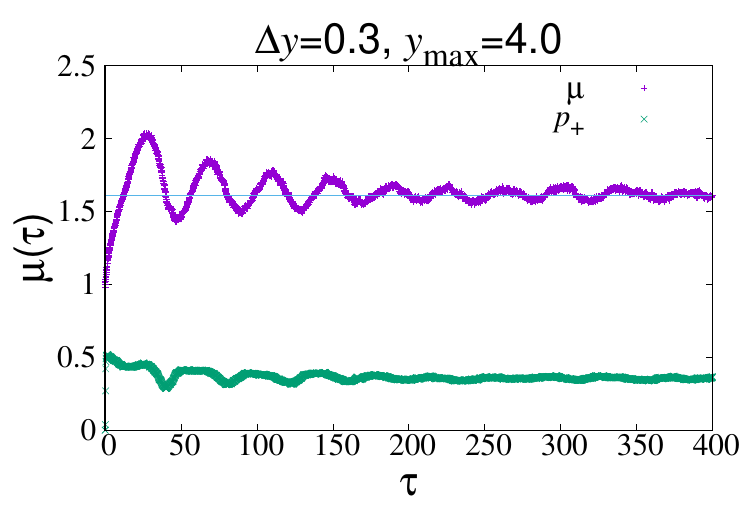}
\includegraphics[width=5.5cm]{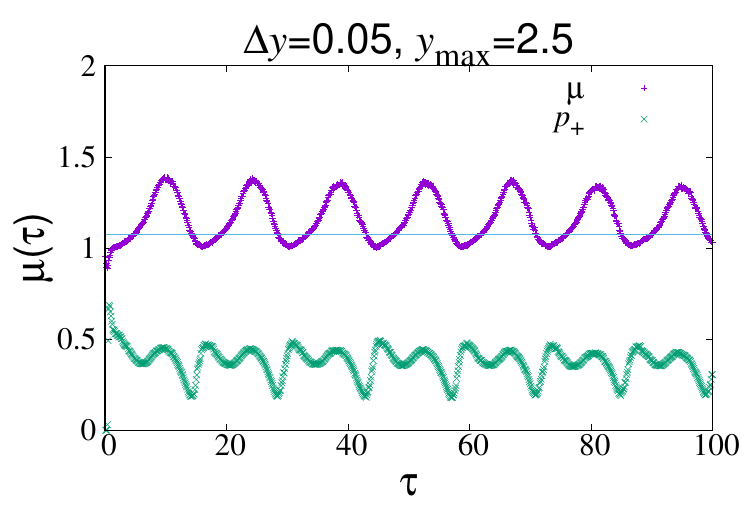}\includegraphics[width=5.5cm]{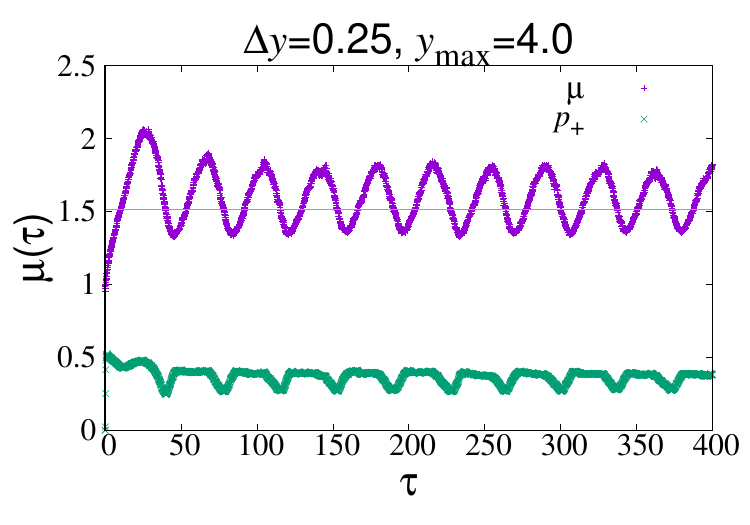}
\includegraphics[width=5.5cm]{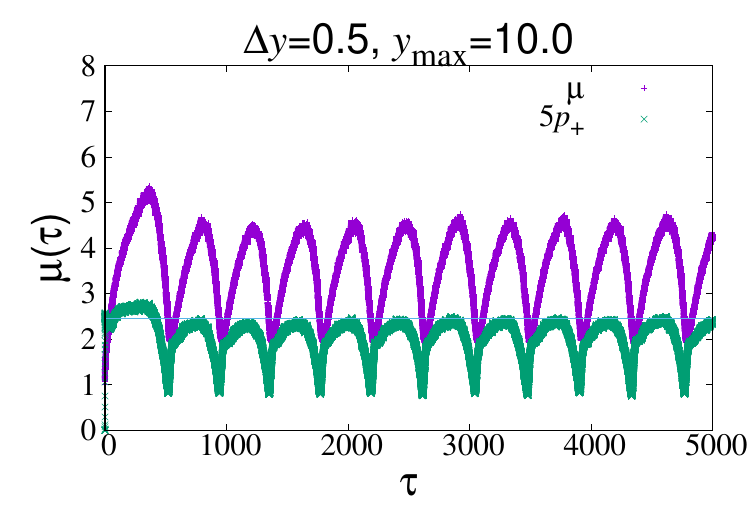}\includegraphics[width=5.5cm]{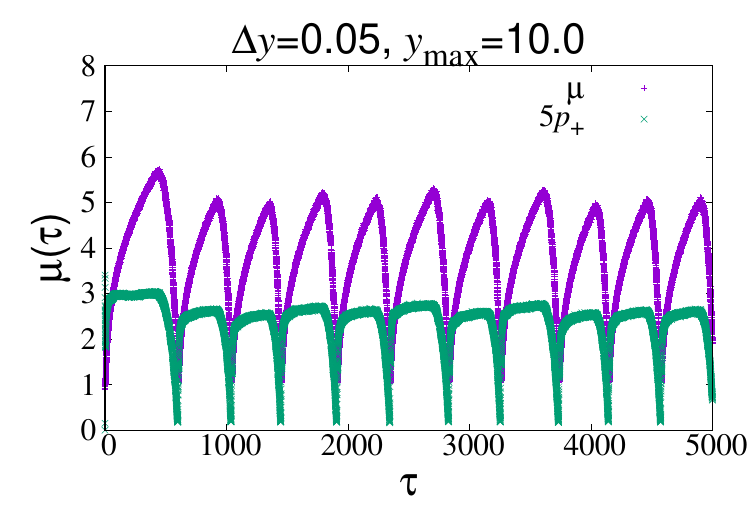}
\includegraphics[width=5.5cm]{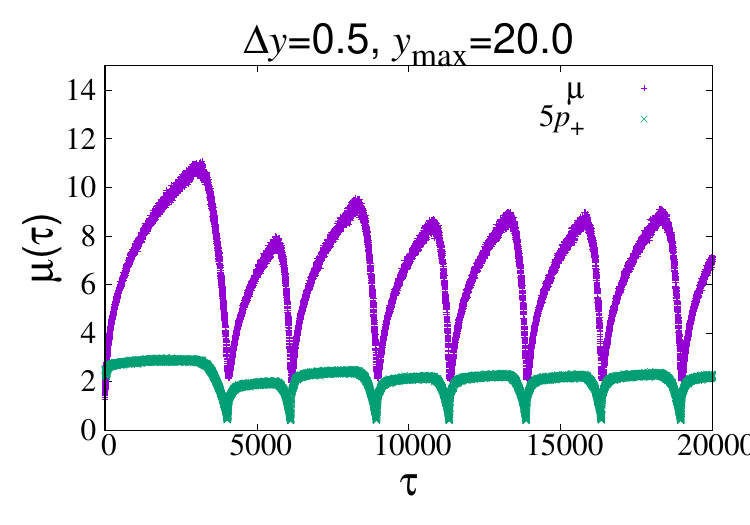}\includegraphics[width=5.5cm]{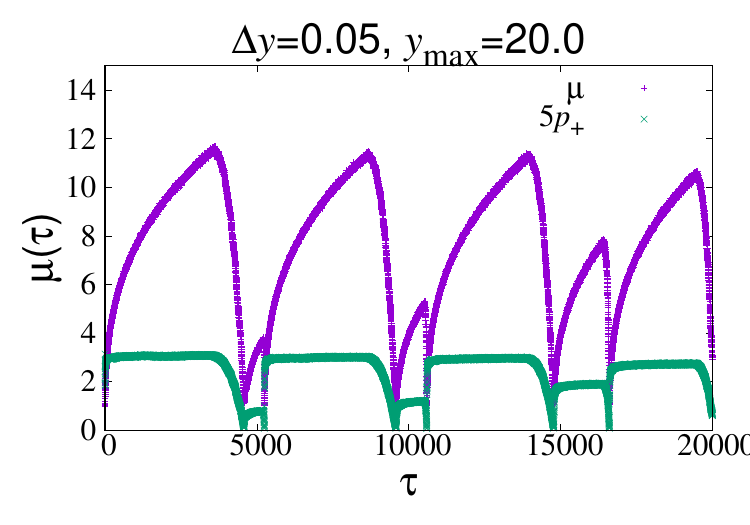}
\end{center}
\caption{
\label{fig: 1}
Numerical evaluation of the dimensionless mean droplet radius $\mu(t)$, for
various choices of the dimensionless initial droplet size dispersion $\Delta y$, 
and cutoff droplet size $y_{\rm max}$. The graphs also plot the probability 
$p_+$ that the dimensionless droplet radius exceeds the critical size for it to grow.   
The first six panels show the transition from a steady state to a limit cycle, for 
$y_{\rm max}=2.5$ and $y_{\rm max}=4.0$. The lower panels illustrate limit 
cycles, and the onset of chaos, at larger values of $y_{\rm max}$. 
}
\end{figure}

Section \ref{sec: 4} discusses the numerical investigation of the steady-state
solution, and of its stability. After the steady-state solution becomes 
unstable, the system approaches a limit cycle, with a periodic oscillation 
of the supersaturation: some examples are shown in figure \ref{fig: 1}. 
Figure \ref{fig: 1} also shows that the model may exhibit chaotic behaviour.
Section \ref{sec: 4} also compares a numerical 
evaluation of the limit cycle with a theory which approximates the 
period, amplitude and profile of the supersaturation fluctuations.
Section \ref{sec: 5} is a conclusion, considering the relevance of the 
results to the classic problem of Ostwald ripening, and to 
atmospheric science.

\section{Equations of motion}
\label{sec: 2}

\subsection{Fundamental equations}
\label{sec: 2.1}

To avoid complicating issues which arise when the precipitating phase is solid, or when 
it has high volume fraction (see, e.g. \cite{Mar+84}), let the physical context be that of 
small liquid droplets dispersed in a gas phase. Gravity effects (treated 
in \cite{Rat+85}) are also neglected.

The equations will be discussed in terms of the atmospheric aerosol, 
which consists of very small water droplets uniformly and randomly
dispersed in air \cite{Mas71}. Let $\Phi$ be the volume fraction of water molecules 
in the air, and $N_0$ be the number of droplets per unit volume. The 
probability density function of the droplet radius $a$ at time $t$ is $p(a,t)$.
The spatial average value of $\Phi$ is $\Phi(t)=\Phi_{\rm e}+\Phi_{\rm s}(t)$, 
where $\Phi_{\rm e}$ is the equilibrium volume fraction and $\Phi_{\rm s}$ is the 
supersaturation. The system is treated in a mean-field approximation, with $\Phi_{\rm s}(t)$ 
independent of position. 

At the surface of small droplets, the equilibrium volume fraction of molecules 
in the gas phase is increased due to the Laplace pressure. If the volume fraction at radius
$r$ from the centre of a droplet is $\Phi(r)$, then
\begin{equation}
\label{eq: 2.1}
\Phi(a) =\Phi_{\rm e}+\frac{\Lambda}{a}
\end{equation}
where ${\Lambda}$, which is proportional to the surface tension $\gamma$,
is known as the capillary length, and $\Phi_{\rm e}$ is the equilibrium 
volume fraction. The capillary length is
\begin{equation}
\label{eq: 2.2}
\Lambda=\frac{2\gamma v_{\rm m}}{kT}\Phi_{\rm e}
\end{equation}
where $v_{\rm m}$ is the molecular volume of water molecules. For water droplets in air, 
$\Lambda\approx 2.1\times 10^{-14}{\rm m}$ (there is no logical difficulty which 
arises from this being less than molecular sizes). 

A droplet of radius $a$ can either grow or shrink depending 
upon whether the supersaturation of the gas surrounding it is greater than or less than 
$\Phi_{\rm cr}=\Lambda/a$. The equation of motion for a growing droplet is determined 
by the rate of evaporation from or condensation onto the droplet surface, as determined
by Fick's law, with diffusion coefficient $D$ (actually, transfer of latent heat is significant
as well \cite{Mas71}, so $D$ is really an effective diffusion 
coefficient $D_{\rm eff}$, representing a weighted 
harmonic mean of molecular and thermal diffusivities).  If the diffusion equation 
describing the volume fraction is treated in a quasi-static approximation then
 \begin{equation}
 \label{eq: 2.3}
\Phi(r)=\frac{[\Phi(a)-\Phi(\infty)]a}{r}+\Phi(\infty)
\end{equation}
 and the rate of change of the droplet radius is
 \begin{equation}
 \label{eq: 2.4}
\frac{{\rm d}a}{{\rm d}t}=D\frac{\partial \Phi}{\partial r}\bigg\vert_{r=a}
\ .
\end{equation}
Settling $\Phi(\infty)=\Phi_{\rm e}+\Phi_{\rm s}$, and using (\ref{eq: 2.3}) to estimate 
the concentration gradient leads to an equation of motion for the droplet sizes:
 \begin{equation}
 \label{eq: 2.5}
\frac{{\rm d}a}{{\rm d}t}=D\left[\frac{\Phi_{\rm s}}{a}-\frac{\Lambda}{a^2}\right]
\ .
\end{equation}
This is the equation of motion obtained by Greenwood \cite{Gre56}.
Because the total quantity of water is fixed, 
\begin{equation}
\label{eq: 2.6}
K=\Phi_{\rm e}+\Phi_{\rm s}(t)+\frac{4\pi}{3V}\sum_j a_j^3
\end{equation}
(where $V$ is the volume of the region) is a conserved quantity.

In the treatment by Lifshitz and Slyozov, where droplet radii grow without 
bound, equation (\ref{eq: 2.5}) is further transformed into an equation of motion for a the ratio
of the droplet radius to a critical radius which is increasing as a function of time. 
In this work, a different path is taken. New droplets are injected into the system 
at a rate $R$, with some specified probability distribution of initial sizes, $p_0(a)$, 
centred on a typical size $a_0$. Droplets either evaporate, or else increase in radius.
Those droplets which grow are removed when their radius reaches $a_{\rm max}$.

\subsection{Equations of motion in dimensionless form} 
\label{sec: 2.2}

It will be convenient to use dimensionless variables. 
Define relative droplet sizes $y_i$, relative supersaturation $x$, 
and a dimensionless time $\tau$ by  
\begin{equation}
\label{eq: 2.7}
y_i(t)=\frac{a_i(t)}{a_0}
\ ,\ \ \ 
x(t)=\Phi_{\rm s}(t)\frac{a_0}{\Lambda}
\ ,\ \ \ 
\tau=\frac{\Lambda Dt}{a_0^3}
\ .
\end{equation}
Also, let $N_0$ be a typical number of droplets present per unit volume.
With these definitions, the equation (\ref{eq: 2.5}) for the droplet radius is
transformed to 
\begin{equation}
\label{eq: 2.8}
\frac{{\rm d}y_i}{{\rm d}\tau}=\frac{x}{y_i}-\frac{1}{y_i^2}
\ .
\end{equation}
Using (\ref{eq: 2.5}), and (\ref{eq: 2.6}), the supersaturation obeys
\begin{equation}
\label{eq: 2.9}
\frac{{\rm d}\Phi_{\rm s}}{{\rm d}\tau}=-\frac{4\pi}{V}\sum_i a_i^2\frac{{\rm d}a_i}{{\rm d}\tau}
=-\frac{4\pi a_0^3}{V}\sum_i y_i^2 \frac{{\rm d}y_i}{{\rm d}\tau} 
=\frac{4\pi a_0^3}{V}\sum_i 1-xy_i
\ .
\end{equation}
The dimensionless supersaturation, $x(\tau)$ then satisfies
\begin{eqnarray}
\label{eq: 2.10}
\frac{{\rm d}x}{{\rm d}\tau}&=&\frac{4\pi a_0^4}{\Lambda V}\sum_i 1-xy_i
\nonumber \\
&=&\alpha f(\tau)\left[1-x\langle y\rangle\right]
\ .
\end{eqnarray}
where $f(\tau)=N(t)/N_0$ is the fraction of droplets present at time $\tau$, relative 
to the reference population density, $N_0$, and where 
\begin{equation}
\label{eq: 2.11}
\alpha=\frac{4\pi N_0a_0^4}{\Lambda}
\end{equation}
is a dimensionless constant, which characterises the initial droplet 
density.

\section{Stiffness and stability}
\label{sec: 3}
 
Now consider an estimate of dimensionless parameter, $\alpha$, defined in (\ref{eq: 2.11}).
Numerical values will be considered for the water-in-air aerosol system (relevant to clouds 
in the terrestrial atmosphere), close to standard atmospheric conditions (sea level, $15^\circ {\rm C})$. 
Some numerical values are: air density: $\rho_{\rm g}=1.26\, {\rm kg\, m}^{-3}$,
water density: $\rho_{\rm l}=1.00\times 10^3\, {\rm kg\,m}^{-3}$, 
surface tension: $\sigma=7.0\times 10^{-2}\, {\rm N\,m}^{-1}$, 
molar volume of water: $V_{\rm m}=10^{-3}/18\,{\rm m}^3$, implying molecular volume
$v_{\rm m}=9.93\times 10^{-29}\,{\rm m}^3$. For the equilibrium volume fraction, 
saturated air at $15\,^\circ {\rm C}$ contains approximately $6.0\,{\rm g\,m}^{-3}$ water vapour, corresponding to a volume of $6.0\times 10^{-6}\ {\rm m}^{-3}$. 
So the volume fraction at equilibrium is $\Phi_{\rm e}=6.0\times 10^{-6}$.

To estimate the value of $\alpha$ for the aerosol system,
assume that $a_0$ takes a typical value for cloud droplets, $a_0=10^{-5}\,{\rm m}$,
and that the liquid water content is $10\%$ of the total water content, so that the liquid 
water content is approximately $\Phi_0=6\times 10^{-7}$. Writing
\begin{equation}
\label{eq: 3.1}
\Phi_0=\frac{4\pi}{3}N_0a_0^3
\end{equation}
leads to $N_0=1.4\times 10^8\,{\rm m}^{-3}$, and hence 
\begin{equation}
\label{eq: 3.2}
\alpha \approx 840
\ .
\end{equation}
The diffusion coefficient was not required for this estimate, but for completeness 
the diffusion coefficient of water molecules in air is approximately  
$D=2.0\times 10^{-5}\,{\rm m}^2{\rm s}^{-1}$. Using this value would neglect effects of cooling 
of evaporating droplets, requiring replacement of latent heat \cite{Mas71}. Strictly 
speaking, a smaller effective diffusion coefficient, $D_{\rm eff}$, should be 
used (but this refinement will not be pursued here).

The parameter $\alpha$ in equation (\ref{eq: 2.10}) has been shown to be large 
for the atmospheric aerosol, and large values will also obtain in other systems 
where Ostwald ripening might occur. This large parameter is indicative of 
(\ref{eq: 2.10}) behaving as a \lq stiff' equation, in the sense that small deviations  
of $x(\tau)$ will be corrected on a very short timescale. This could imply that the numerical 
solution might be challenging, requiring a very short timestep, or it might be associated 
with an inherent instability, with ${\rm d}x/{\rm d}\tau$ exhibiting rapid and possibly 
chaotic fluctuations. 

If ${\rm d}x/{\rm d}\tau$ is a well-behaved function of the dimensionless time $\tau$, 
then in the limit as $\alpha\to \infty$ the distribution of values of $y$ is constrained:
\begin{equation}
\label{eq: 3.3}
\lim_{\alpha \to \infty}x\langle y\rangle=1
\ .
\end{equation}

It has been argued that the dimensionless supersaturation $x(\tau)$ and
the dimensionless mean droplet radius $\mu(\tau)\equiv \langle y \rangle$
are related by $x(\tau)\mu(\tau)=1$.
Let us assume that the system is in a steady state, where 
$\mu(\tau)=\mu_0$, and $x(\tau)=x_0$, with $\mu_0 x_0=1$. 
If this equilibrium is perturbed by a small fluctuation of the supersaturation, $\delta x(\tau)$, 
there will be a time-dependent fluctuation of the mean. To leading 
order, this response is linear, described by a kernel $K(\Delta \tau)$:  
\begin{equation}
\label{eq: 3.4}
\delta \mu(\tau)=\int_{-\infty}^\tau {\rm d}\tau'\ K(\tau-\tau')\,\delta x(\tau')
\ .
\end{equation}
If equation (\ref{eq: 3.3}) is to be valid for all time, we must have
$x_0 \delta \mu(\tau)+\mu_0 \delta x(\tau)=0$, so that 
\begin{equation}
\label{eq: 3.5}
\delta x (\tau)=-x_0^2\int_{-\infty}^\tau {\rm d}\tau'\ K(\tau-\tau')\,\delta x(\tau')
\ .
\end{equation}
We must assume that this relation has been satisfied at all times up to 
$\tau$, so that the unknown value of $\delta x(\tau)$ is expressed 
in terms of the known values of $\delta x(\tau')$ with $\tau'<\tau$.

Consider the criterion for $\delta x(\tau)$ to remain bounded. 
Because (\ref{eq: 3.5}) is a linear relationship, we can conclude 
that the equation is unstable if there is a growing solution for any 
complex frequency. If, for some small constant $\epsilon$,
\begin{equation}
\label{eq: 3.6}
\delta x(\tau)=\epsilon \exp(z\tau)
\ ,\ \ \ 
z=\lambda+{\rm i}\omega
\end{equation}
the stability is then determined the zero of 
\begin{equation}
\label{eq: 3.7}
F(z)=1+x_0^2\int_0^\infty {\rm d}t\ K(t)\exp(-zt)
\end{equation}
with the largest value of ${\rm Re}(z)$. If this is positive, the
fixed point is unstable. 

\section{Simulations}
\label{sec: 4}

\subsection{Simulations of evolution of model}
\label{sec: 4.1}

Equations (\ref{eq: 2.8}) and (\ref{eq: 2.10}) were integrated numerically, 
with (\ref{eq: 2.10}) written in the form
\begin{equation}
\label{eq: 4.1}
\frac{{\rm d}x}{{\rm d}\tau}=\frac{A}{R}\sum_i 1-xy_i
\end{equation}
with $A$ a positive constant and $R$ the rate of injection of new droplets.
The simulations are regarded as representing the droplets in a unit volume 
of fluid (so that $V=1$ in equation (\ref{eq: 2.10})). One droplet is added 
per timestep $\delta \tau$, so that $R=1/\delta \tau$. 
The distribution of injected particle sizes, $y_0$, was a truncated 
Gaussian, 
\begin{equation}
\label{eq: 4.2}
p_0(y_0)=\left\{ \begin{array}{cc}
C\exp\left[-\frac{(y-y_{\rm c})^2}{2\Delta y^2}\right] & y_0\ge 0 \cr  
0        & y_0<0\ ,\ \ y_0>y_{\rm max} 
\end{array}
\right.
\end{equation}
with mean $y_{\rm c}=1$, where $C$ is a normalisation constant.
The simulation determined the mean value $\langle N\rangle$ of the number of droplets, 
after a \lq warmup' time $\tau_{\rm w}$, so that the parameter $\alpha$ in equation
(\ref{eq: 2.10}) is
\begin{equation}
\label{eq: 4.3}
\alpha=A\langle N\rangle \delta \tau
\ .
\end{equation}
All of the simulations reported here used $A=25.0$ and $\delta \tau=10^{-3}$.
Numerical tests showed that the results are insensitive to varying these parameters.

The time evolution was investigated for a different choices of the 
standard deviation of the injection size distribution, $\Delta y$, 
and the upper cutoff of droplet size, $y_{\rm max}$.
Results are shown in figure \ref{fig: 1}, showing plots of the mean 
droplet radius, $\mu(\tau)$.
The plots also display the quantity $p_+(\tau)$, 
which is the probability that a droplet has size greater than $1/x(\tau)$, 
as a function of time. This quantity is the fraction of droplets which are growing
at any instant of time. It is required for the considerations 
in section \ref{sec: 4.3} below. 
The mean value of the quantity
\begin{equation}
\label{eq: 4.2}
\Delta(\tau)\equiv |1-x(\tau)\mu(\tau)|
\end{equation}
was also evaluated, averaged over times $\tau$ which are greater than $\tau_{\rm w}=1$. 
In all of the simulations $\alpha\gg 1$, and 
$\Delta (\tau)$ is very small implying that $x\mu\sim 1$, in accord with 
(\ref{eq: 3.3}). The parameters for all of the simulations illustrated 
in figure \ref{fig: 1} are listed in table \ref{tab: 1}.

For some choices of $(\Delta y,y_{\rm max})$, the values of $\mu(\tau)$,  
and correspondingly the supersaturation $x(\tau)$, converge to constant values 
as $\tau\to \infty$.  For other choices the mean value $\mu(\tau)$, 
approaches a limit cycle, or may become chaotic. The period of the limit cycle of $\mu(\tau)$ varies 
over orders of magnitude.

\subsection{The steady-state solution and its stability}
\label{sec: 4.2}

The steady-state solution was investigated by modifying the code 
to evolve the droplet radii with a fixed value of the supersaturation, $x_0$.
This code determines the mean droplet radius in the long-time limit, 
$\mu_0$, as a function of $x_0$. 
In equilibrium, the steady-state saturation is determined by solving the equation
$x_0\mu_0(x_0)=1$. Values of $\mu_0$ are tabulated in table \ref{tab: 1} and
are also shown as lines on each of the plots in figure \ref{fig: 1}. In the cases where the 
evolution is stable, the long-time limit of $\mu_0(\tau)$ agrees with $\mu_0$. 

\begin{figure}
\begin{center}
\includegraphics[width=6.0cm]{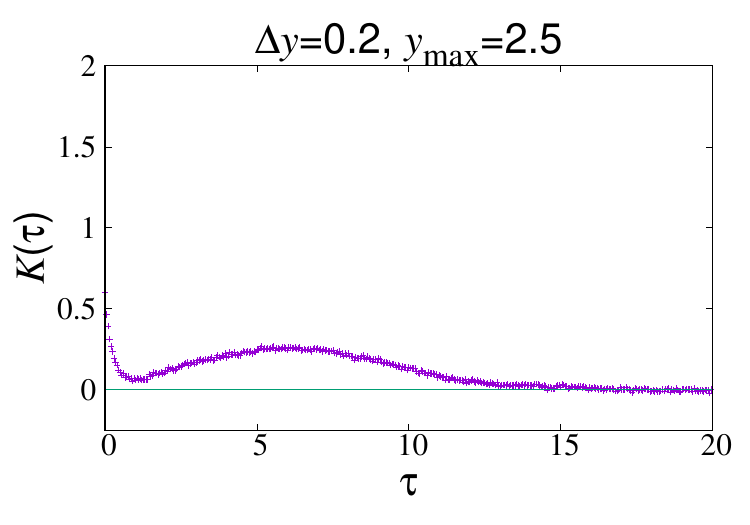}\includegraphics[width=6.0cm]{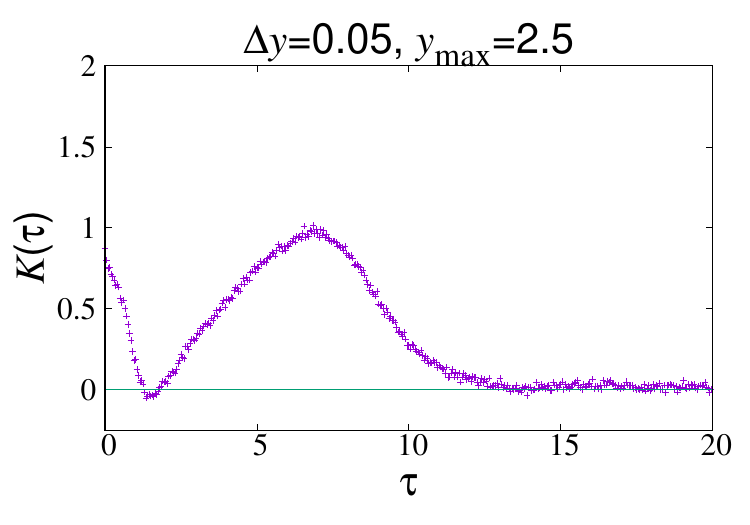}
\includegraphics[width=6.0cm]{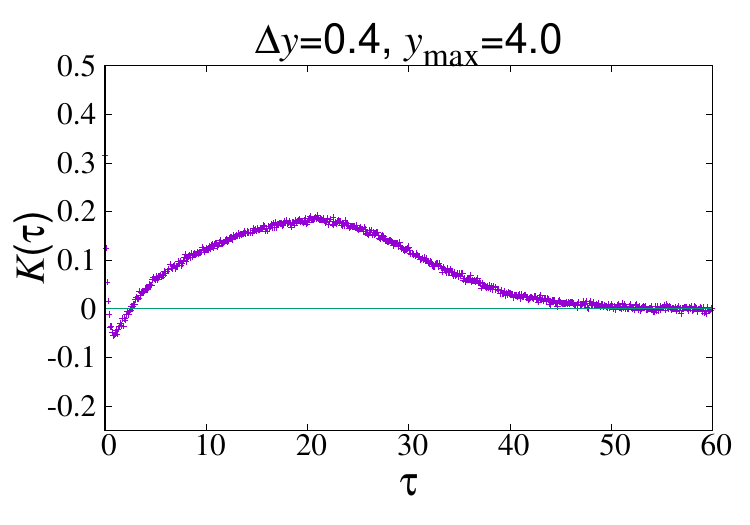}\includegraphics[width=6.0cm]{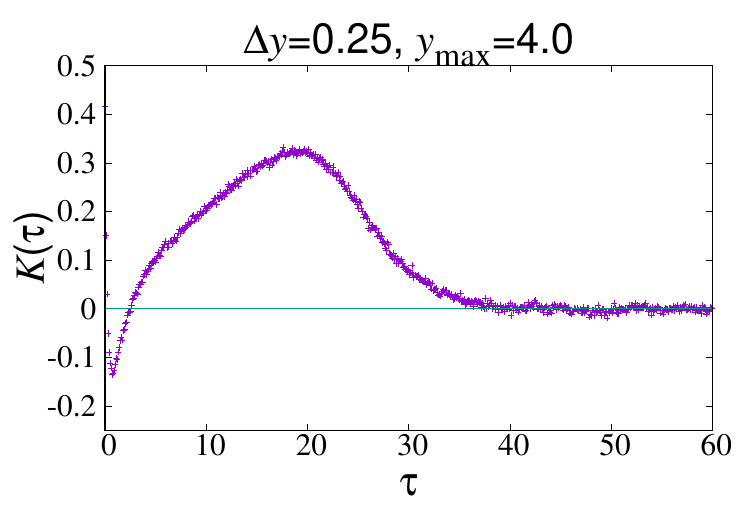}
\end{center}
\caption{
\label{fig: 2}
Numerical evaluation of the response kernel $K(\tau)$, for
the some of the choices of the initial droplet size dispersion $\Delta y$, 
and cutoff droplet size $y_{\rm max}$ in figure \ref{fig: 1}.
}
\end{figure}

In order to assess the stability of the steady-state solution, it is necessary
to evaluate the response kernel $K(\Delta \tau)$, defined by equation (\ref{eq: 3.4}).
The kernel was approximated numerically by determining the mean value $\tilde \mu(\tau)$, 
which results from $x(\tau)$ having the following specified form:
\begin{equation}
\label{eq: 4.3}
\tilde x(\tau)=\left\{
\begin{array}{cc}
x_0 & -\tau_{\rm min}<\tau \le 0 \cr
x_0+\Delta x & 0<\tau<\Delta \tau \cr
x_0 & \Delta \tau\le \tau_{\rm max} \cr
\end{array}
\right.
\ .
\end{equation}
The kernel is then approximated by 
\begin{equation}
\label{eq: 4.4}
K(\tau)=\frac{\tilde \mu(\tau)-\mu_0}{\Delta x\Delta \tau}
\end{equation}
with $\Delta x$ and $\Delta \tau$ made as small as possible, consistent 
with obtaining stable numerical results. Also, $\tau_{\rm min}$ and $\tau_{\rm max}$
are made large enough to yield stable results. The numerically determined response kernels for some
of the examples in figure \ref{fig: 1} are plotted in figure \ref{fig: 2} (the data for figure \ref{fig: 2} used 
$\Delta t=0.1$ and $\Delta x=0.05$).
Table \ref{tab: 1} lists the location of the zero of the response function 
$F(z)$ (defined by (\ref{eq: 3.7})) with the largest real part for various cases, together with their 
other parameters. These values are consistent with the real part of the zero changing sign 
at the onset of instability. Table \ref{tab: 1} also lists the period $T$ and peak-to-trough 
amplitude $\Delta \mu$ of the oscillations of $\mu(\tau)$. The period $T$ and the frequency 
of the zero of the response function, $\omega_0$, can be seen to be related
(approximately) by $\omega T\approx 2\pi$. 

\begin{table}[ht]
\caption{Tabulation of parameters for the examples in figure \ref{fig: 1}. Some values 
are absent because of limitations of the numerical method.}
\centering
\begin{tabular}{c c c c c c c}
\hline\hline
 $(\Delta y,y_{\rm max})$&$\mu_0$&$\alpha $& $\langle \Delta (\tau)\rangle $ 
 & $(\lambda,\omega)$&$\langle p_+\rangle$&$(T,\Delta \mu)$ \\ [0.5ex]
\hline
\\ [ 1 ex]
$(0.2,2.5)$   & $1.290$   &  $49.2$ &  $8.8\times 10^{-4}$ & $(-0.124,0.415)$   & $0.425$ &${\rm stable}$ \\ 
$(0.1,2.5)$ & $1.149$   &  $50.6$ &  $9.9\times 10^{-4}$ & $(-0.005,0.488)$       & $0.360$ &${\rm stable}$ \\
$(0.05,2.5)$   & $1.075$   & $52.7 $ &  $11.2\times 10^{-4}$ & $(0.086,0.518)$       & $0.376$ & $(14.5,0.29)$ \\ 
$(0.4,4.0)$ & $1.794$ & $53.5$ &  $7.2\times 10^{-4}$ & $(-0.026,0.136)$ & $0.401$ & ${\rm stable}$ \\
$(0.3,4.0)$ & $1.614$   & $44.7$ &  $9.8\times 10^{-4}$ & $(-0.006,0.166)$       & $0.368$ &${\rm stable}$  \\
$(0.25,4.0)$  & $1.519$   & $41.5$ &  $11.2\times 10^{-4}$ & $(0.014,0.180)$       & $0.368$ &$(36.7,0.42)$  \\   
$(0.5,10.0)$   & $2.453$   & $43.4 $ &  $7.4\times 10^{-4}$ & $-$       & $0.410$ & $(367,2.25)$ \\ 
$(0.05,10.0)$ & $-$ & $22.6 $ &  $19.6\times 10^{-4}$ & $-$ & $0.477$ & $(478,5.09)$ \\
$(0.5,20.0)$ & $-$   & $34.2$ &  $7.5\times 10^{-4}$ & $-$       & $0.427$ &$(2500,6.45)$  \\
$(0.05,20.0)$  & $-$   & $20.9$ &  $16.7\times 10^{-4}$ & $-$       & $0.502$ &${\rm chaotic}$  \\   
\hline
\end{tabular}
\label{tab: 1}
\end{table}

\begin{figure}
	\includegraphics[width=10.0cm]{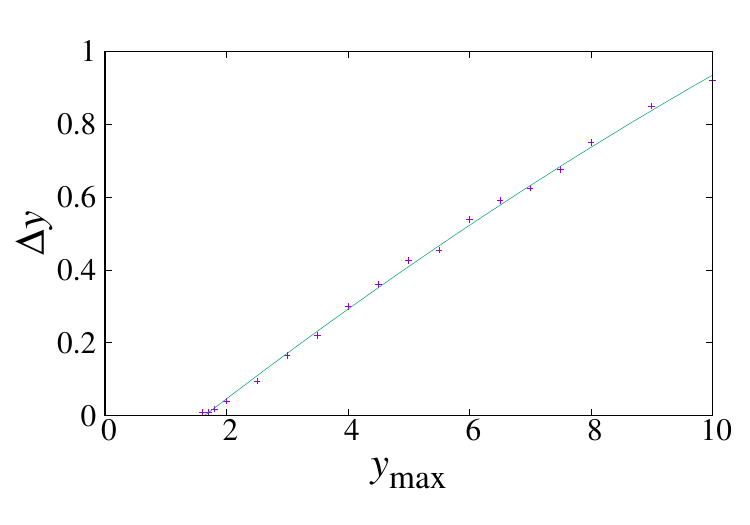}
\caption{
\label{fig: 3}
Boundary of the unstable region in the $(\Delta y,y_{\rm max})$ plane 
(the unstable region lies to the left of the line). The smooth curve, 
$\Delta y=0.216+0.135 y_{\rm max}-0.002 y_{\rm max}^2$ is a guide to the eye.
}
\end{figure}

Figure \ref{fig: 3} shows the border of the stable region in the $(\Delta y,y_{\rm max})$ 
plane. Because evaluation of the kernel is numerically intensive, stability was assessed by plotting $\mu(\tau)$, 
rather than by seeking roots of (\ref{eq: 3.7}). 

\subsection{Form of the limit cycle}
\label{sec: 4.3}

As $y_{\rm max}$ increases, both the amplitude $\Delta \mu$ and the period $T$
of the limit-cycle oscillations increase rapidly. The oscillations of $\mu(t)$ 
also develops a distinctive asymmetric profile. These features can be given a 
quantitative explanation, which is closely related to the standard model for 
Ostwald ripening. There are, however, significant points of difference. 

In the standard Ostwald ripening scenario, no additional droplets are injected and the 
mean radius of the droplets is argued to have an asymptotic growth of the form
\cite{Lif+61}
\begin{equation}
\label{eq: 4.5}
\langle a\rangle \sim \left(\frac{4}{9}Dt\right)^{1/3}
\ .
\end{equation}
On the basis of (\ref{eq: 4.5}), we might guess that, when $y_{\rm max}\gg 1$, 
the mean size of the droplet distribution may grow as $\langle y\rangle\sim (4\tau /9)^{1/3}$, 
until it reaches $y_{\rm max}$. This suggests that the period would be 
$T\sim 729 y_{\rm max}^3/64$. This is an over-estimate of the period. 

A more persuasive model can be developed as follows. Let $\Delta \tau$ be the 
time since the last \lq reset' event, at which $\mu(\tau)$ goes to its lowest value
in the limit cycle.
When $y_{\rm max}\gg 1$, the PDF of the dimensionless droplet radius $y$ 
develops two peaks. There is a population of larger droplets, centred on $y_{\rm p}(\tau)$, 
which increase in size until they reach $y_{\rm max}$, and a population of droplets which decrease 
in size until they evaporate.
As the larger droplets grow, the degree of supersaturation decreases. The suppressed supersaturation 
implies that the droplets which are being injected evaporate in an environment of low supersaturation.
When the droplets of radius $y_{\rm p}$ reach radius $y_{\rm max}$, they are abruptly 
removed from the system, the supersaturation rebounds, and $\mu(\tau)$ returns to its minimum value. 

Assume that the distribution $p(y,\tau)$ evolves so that, shortly after a \lq reset' has occurred, the 
probability that $y>1/x$ reaches a value $p_+$, which subsequently remains approximately 
constant, until the next reset event. While the value of $p_+$ is difficult to estimate theoretically, 
numerical values for $p_+$ are readily obtained: see figure \ref{fig: 1} and table \ref{tab: 1}.

According to this model, the mean value of the droplet radius is approximated by
\begin{equation}
\label{eq: 4.6}
\mu(\Delta \tau) \sim \mu_{\rm e}(1-p_+)+p_+y_{\rm p}(\Delta \tau) 
\end{equation}
where $\mu_{\rm e}$ is the mean dimensionless droplet radius for droplets 
which are injected with unit radius, and which evaporate in an environment 
with negligible supersaturation. Because the time-dependence of the radius 
of such a droplet is $y(\tau)\sim 1-(3\tau)^{1/3}$, we find $\mu_{\rm e}=3/4$. 
   
The equation of motion for $y_{\rm p}$ is
\begin{equation}
\label{eq: 4.7}
\frac{{\rm d}y_{\rm p}}{{\rm d}\tau}=\frac{1}{y_{\rm p}^2}
\left[\frac{y_{\rm p}}{\mu(\Delta \tau)}-1\right]
\end{equation}
When $y_{\rm p}\gg 1$, (\ref{eq: 4.6}) can be approximated by
$\mu\sim p_+ y_{\rm p}$, so that (\ref{eq: 4.7}) is approximated by
\begin{equation}
\label{eq: 4.8}
\frac{{\rm d}y_{\rm p}}{{\rm d}\tau}=\frac{1}{y_{\rm p}^2}\left[{\frac{1}{p_+}-1}\right]
\ .
\end{equation}
The solution is
\begin{equation}
\label{eq: 4.9}
y_{\rm p}(\Delta \tau)\sim \left[\frac{3(1-p_+)}{p_+}\Delta \tau \right]^{1/3}
\end{equation}
where $\Delta \tau$ is the time since the minimum of the cycle. This solution is valid 
until $y_{\rm p}$ reaches $y_{\rm max}$, at which point 
the reset occurs. This latter condition defines the period. 
In the limit as $y_{\rm max}\to \infty$, we can ignore, the displacement
$\Delta \tau_0$ can be neglected, and the 
period and amplitude are then approximated by 
\begin{equation}
\label{eq: 4.10}
T\sim \frac{p_+}{3(1-p_+)}y_{\rm max}^3
\ ,\ \ \ 
\Delta \mu \sim p_+ y_{\rm max}
\ .
\end{equation}
The values of $p_+$ listed in table \ref{tab: 1} are the average values of this parameter, after
a \lq warmup' time equal to twice the predicted period.

\begin{figure}
	\includegraphics[width=10.0cm]{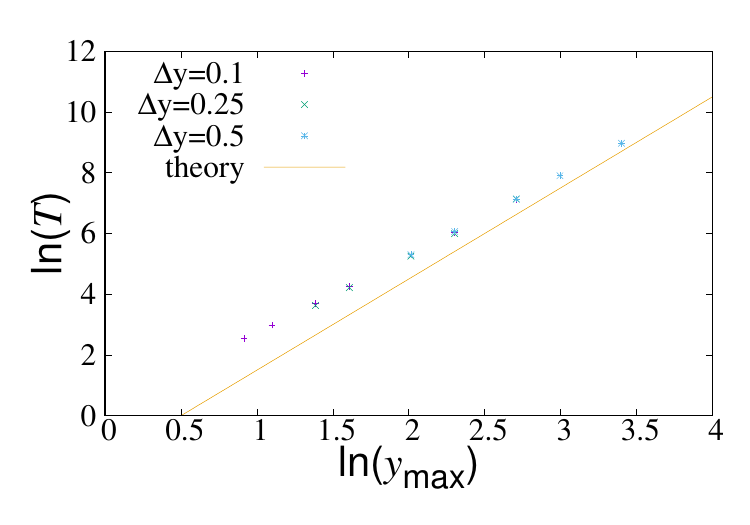}
\caption{
\label{fig: 5}
Predicted versus measured period $T$, as a function of $y_{\rm max}$, 
for different values of $\Delta y$. The theoretical line uses $p_+=0.4$ 
in equation (\ref{eq: 4.10}).
}
\end{figure}

\begin{figure}
	\includegraphics[width=10.0cm]{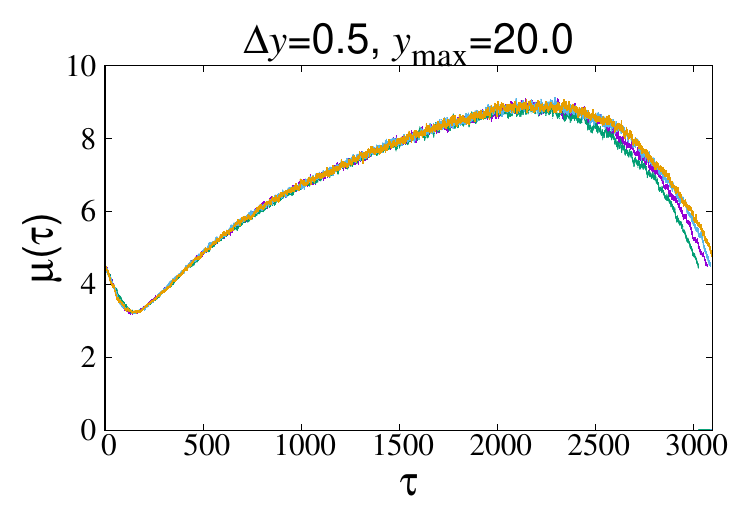}
\caption{
\label{fig: 6}
Illustrating the form of the limit cycle, for the case $(\Delta y,y_{\rm max})=(0.5,20.0)$. 
The evolution of $\mu(t)$, starting from a local minimum, is overlaid for four different 
cycles. These curves overlay each other quite precisely, 
consistent with the instability leading to a simple limit-cycle, without period-doubling
or chaos.}
\end{figure}

Figure \ref{fig: 5} shows the comparison between the measured period $T$ 
and the theoretical value given by equation (\ref{eq: 4.10}), for three different values
of $\Delta y$. There is fair agreement with (\ref{eq: 4.10}), improving as $y_{\rm max}$ increases. 
The dependence of $T$ upon $\Delta y$ is very weak. 
Figure \ref{fig: 6} overlays $\mu(\tau)$ curves for different cycles, in the case $\Delta y=0.5$, $y_{\rm max}=20$. 
The curves overlap quite precisely, indicating that the instability
leads to a limit cycle, rather than to chaos or period doubling. For smaller $\Delta y$, however, the system
appears to behave chaotically, as is evident from the final panel of figure \ref{fig: 1}. 
The nature of this transition to chaos was not understood.

\section{Concluding remarks}
\label{sec: 5}

This paper has discussed the equation of motion for the supersaturation 
in an Ostwald ripening process. The equation contains a dimensionless 
parameter $\alpha$ which was shown to be large, at least in the case of the atmospheric aerosol.
In section \ref{sec: 3} it was argued that this implies that the product of the mean 
droplet radius and the supersaturation is very close to unity: $x\langle y\rangle=1$. 
This conclusion corresponds to the prediction from a treatment of Ostwald ripening 
by Penrose \cite{Pen97}, who described the macroscopic process of 
Ostwald ripening as a limiting case of the microscopic Becker-D\"oring process,
and is also consistent with the supersaturation field containing a negligible 
of the species which forms the droplet phase. 
The approach in this paper complements that earlier work by showing that 
the constraint  $x\langle y\rangle=1$ may be obtained directly from macroscopic considerations. 

The constraint  $x\langle y\rangle=1$ may be a source of instability in the dynamics of 
Ostwald ripening processes. The mechanism for the instability 
is that the mean droplet size $\langle y\rangle $ depends upon the history of the supersaturation 
$x(t)$. As the supersaturation $x(t)$ is continually adjusted so that the constraint $x\langle y\rangle$
is satisfied, the evolution of $x(t)$ may become unstable. 
This paper analysed a version of the Ostwald ripening process in which new droplets are continually 
injected, and large droplets removed. This model is attractive for studying instability because it 
can exhibit a steady-state. The criterion for instability depends zeros of a response function $F(z)$,
constructed from the the Laplace transform of the response kernel, equation (\ref{eq: 3.7}). 
It was shown that the model can indeed have an instability, with a threshold which is correctly 
predicted by zeros of $F(z)$. 
These observations suggest that the classic Ostwald ripening process may also have 
instabilities. These would be harder to analyse, because there is not a true steady state 
condition about which the growth of perturbations can be analysed.

Understanding the growth of raindrops from microscopic  water droplets in clouds is a 
challenging problem, because the rate of collisions between droplets is very low. While 
other mechanisms, such as the \lq lucky droplet' concept \cite{Wil23} may help to resolve 
these difficulties, it is desirable to fully explore non-collisional mechanisms for droplet
growth in the atmospheric aerosol. The unstable phase of the model which is analysed 
in this paper exhibits periodic precipitation events, and could be relevant to atmospheric 
precipitation \cite{Mas71}, or to the periodic precipitation phenomena which are observed when 
a system is slowly driven through a critical point of phase separation \cite{Vol+97}.

\end{document}